\documentclass[%
 reprint,
 amsmath,amssymb,
 aps,
]{revtex4-2}

\usepackage{graphicx}
\usepackage{dcolumn}
\usepackage{bm}
\usepackage{hyperref}
\usepackage[mathlines]{lineno}

\begin{document}

\preprint{APS/123-QED}

\title{A minimalist approach to BL Lacertae: explaining gamma-ray spectral and temporal variability with a single physical parameter}

\author{Raniere de Menezes}
 \email{Contact author: raniere@cbpf.br}
 \altaffiliation{Centro Brasileiro de Pesquisas F\'isicas, 22290-180, Rio de Janeiro, RJ, Brazil}
\affiliation{%
 Centro Brasileiro de Pesquisas F\'isicas, 22290-180, Rio de Janeiro, RJ, Brazil
}%

\author{Francesco Massaro}
\author{Elisa Visentin}
\affiliation{
 INFN -- Istituto Nazionale di Fisica Nucleare, Sezione di Torino, via Pietro Giuria 1, I-10125 Turin, Italy
}%
\affiliation{
 Dipartimento di Fisica, Universit\`a degli Studi di Torino, via Pietro Giuria 1, I-10125 Torino, Italy
}%
\author{Federico Di Pierro}
\affiliation{%
 INFN -- Istituto Nazionale di Fisica Nucleare, Sezione di Torino, via Pietro Giuria 1, I-10125 Turin, Italy
}%
\author{Haocheng Zhang}
\affiliation{%
 University of Maryland Baltimore County, Baltimore, MD 21250, USA
}%
\affiliation{%
 NASA Goddard Space Flight Center, Greenbelt, MD 20771, USA
}%

\date{\today}

\begin{abstract}
The eponymous BL Lac object BL Lacertae is one of the most well-monitored active galactic nuclei, frequently observed from radio to gamma rays. Its relatively soft $\gamma$-ray spectrum peaks near 500~MeV, and since 2020 it has undergone an exceptional series of flaring episodes. The observed emission is well described by synchrotron self-Compton (SSC) models, with negligible contribution from external seed photons.
We investigate the physical origin of BL~Lacertae's $\gamma$-ray temporal and spectral variability using data from the Large Area Telescope (LAT) on board the \textit{Fermi} Gamma-ray Space Telescope, and show that this variability can be explained by a single varying parameter, namely the electrons' peak energy, $\gamma_p$, under a single-zone SSC scenario with a log-parabolic electron distribution.
We use a Markov chain Monte Carlo to estimate the spectral parameters of BL Lacertae over time, selected from an adaptive-binned gamma-ray light curve. We then study the correlation between the inverse Compton peak luminosity, $L_{IC}$, and the position of this peak on the SED energy axis, $E_p$, and compare it with what is expected for a single-zone SSC scenario when only one parameter is free to vary.
We find a correlation $L_{IC} = 10^{42.33\pm0.15\pm0.18_{sys}}E_p^{0.98\pm0.05\pm0.06_{sys}}$, consistent, within the errors, with the linear relation $L_{\mathrm{IC}} \propto E_p$, expected when $\gamma_p$ is the only free parameter in the assumed SSC model. This result supports a minimalist SSC scenario in which changes in $\gamma_p$ dominate the observed temporal and spectral variability of BL~Lacertae.
\end{abstract}

\maketitle


\section{Introduction}
\label{sec:intro}

BL Lac objects are characterized by the lack of strong optical–infrared lines \citep[equivalent widths $< 5~\rm{\AA}$;][]{stickel1991complete,deMenezes2019characterization,deMenezes2020optical_camp_X}, significant variability across the electromagnetic spectrum on time scales from years to minutes \citep{albert2007variable,aharonian2007exceptional,arlen2012rapid}, and a non-thermal spectral energy distribution (SED) with a synchrotron bump peaking between the infrared and X-rays \citep{demenezes2025oringin_of_IR_colors} and a high-energy bump in the gamma-ray band \citep{tavecchio1998constraints,bottcher2013leptonic}.

The highly variable gamma-ray emission from BL Lac objects likely originates in a compact region--unresolved by current instruments--located at distances ranging from subparsec scales to a few parsecs from the central black hole \citep{agudo2011location,bottcher2016gamma}. Flaring states usually come with changes in the hardening of the gamma-ray spectrum \citep{deng2025spectral}, and understanding how the flux correlates with the spectral hardening can reveal the underlying mechanism behind the observed variability \citep{massaro2008_variability}.

In this work, we analyze nearly 17 years of \textit{Fermi} Large Area Telescope \citep[LAT;][]{atwood2009LAT} observations to investigate the spectral variability of the eponymous BL Lac object BL~Lacertae in gamma rays. Our choice of this target is motivated by the availability of extensive LAT data, allowing the construction of a large sample of high-quality, log-parabolic gamma-ray spectra across different activity states. In the LAT Fourth Source Catalog\footnote{\url{https://fermi.gsfc.nasa.gov/ssc/data/access/lat/14yr_catalog/}} \citep[4FGL-DR4;][]{abdollahi2020_4FGL, abdollahi2022_4FGL_DR3}, BL~Lacertae is the second most significant BL~Lac object (after Mrk~421), with its spectral peak located near 500~MeV. Mrk~421 is not considered here, as its gamma-ray emission peaks around 100~GeV, requiring joint analyses with imaging atmospheric Cherenkov telescopes (IACTs) to precisely constrain its spectral shape across different activity states. We propose that the temporal and spectral variability of BL~Lacertae can be interpreted through a single physical parameter, assuming its gamma-ray emission arises from a one-zone synchrotron self-Compton (SSC) model \citep{maraschi1992jet,bloom1996analysis} and that the electron energy distribution follows a log-parabolic form \citep{massaro2004logpar1,massaro2004logpar2,massaro2006logpar3,dermer2015near} $n(\gamma) = n_0 \left(\gamma/\gamma_0\right)^{-s - r\log_{10}(\gamma/\gamma_0)}$, where $n_0$ is the density of emitters per interval of $\gamma$ (cm$^{-3}$), $\gamma_0$ is the electrons' turn-over energy, $r$ the spectral curvature, and $s$ the spectral slope. This simplified model has been repeatedly applied to BL~Lacertae, consistently providing a good description of the data \citep{ravasio2002bl,ravasio2003bepposax,abdo2011first}.

\begin{figure*}
    \centering
    \includegraphics[width=\linewidth]{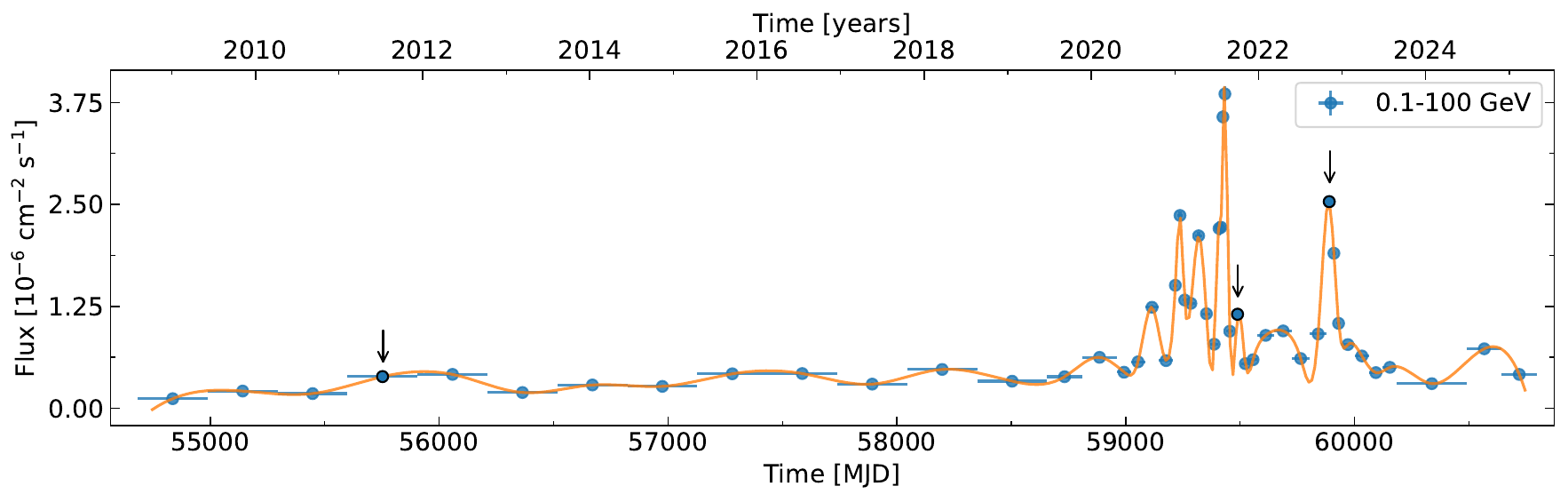}
    \includegraphics[width=\linewidth]{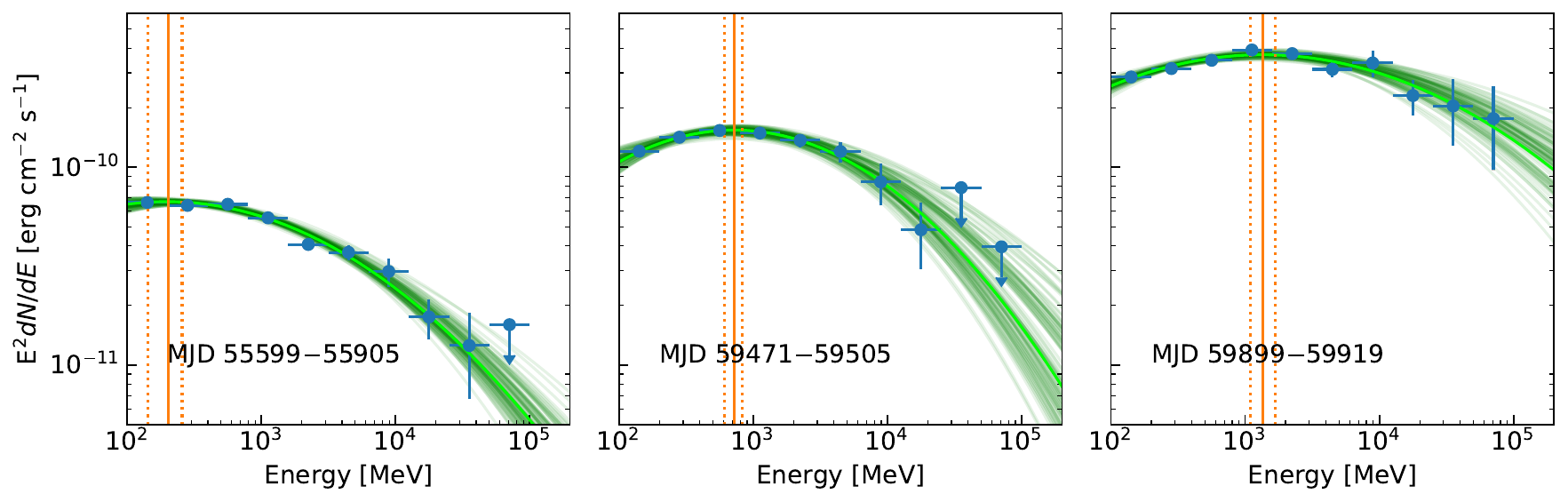}
    \caption{Temporal (top) and spectral (bottom) evolution of BL Lacertae in gamma rays. In the integrated photon-flux light curve shown in the top panel, the three arrows mark the time bins corresponding to the three spectra shown in the bottom panels. The chosen TS for upper limits is set to 25. In the bottom panels, we can see that the log-parabola (green solid line) maximum likelihood spectral peak (orange vertical line) shifts towards higher energies when the gamma-ray flux increases. The dotted vertical lines represent the $68\%$ containment interval for the MCMC posterior distribution on $E_p$. The spectra presented here are in the observed frame.}
    \label{fig:LCs_and_specs}
\end{figure*}

Our main goal is to study how the isotropic inverse Compton (IC) peak luminosity, $L_{IC}$, evolves with the peak energy in the SED, $E_p$. If the gamma-ray emission of BL~Lacertae is dominated by Thomson scattering \citep[see][]{massaro2008_variability}, then $L_{IC} \propto E_p$ when the only varying parameter is the energy at the peak of the electron distribution $\gamma_{p}$ (note that $\gamma_{p}\propto\gamma_0$ when given a fixed $r$ and $s$, i.e., $\gamma_{p} = \gamma_010^{-s/2r}$)\footnote{We can also write $\gamma_p$ in terms of the peak of the $\gamma^3n(\gamma)$ distribution $\gamma_{3p} = \gamma_p10^{-3/2r}$, which is frequently used in blazar literature \citep[see, e.g., Sect. 4.2 in][]{Tramacere2011jetset}.}, $L_{IC} \propto E_p^2$ when only the magnetic field $B$ varies, and $L_{IC} \propto E_p^4$ when only the beaming factor $\delta$ varies. In contrast, if Klein–Nishina effects dominate, the luminosity must scale as $L_{IC} \propto E_p^4$, with $\delta$ as the sole varying parameter.

This paper is organized as follows. Section~\ref{sec:data_selection} presents the gamma-ray LAT data selection and analysis. In Section~\ref{sec:results}, we report our findings and show that the spectral variability can be explained by changes in a single SSC parameter, namely the energy at the peak of the electron distribution. In Section~\ref{sec:discussion_and_conclusions}, we discuss and summarize our results. Throughout this work we adopt $H_0 = 0.70$ km s$^{-1}$ Mpc$^{-1}$, $\Omega_m = 0.30$, and $\Omega_{\Lambda} = 0.70$ \citep{tegmark2004cosmological}.



\section{Data selection and analysis}
\label{sec:data_selection}

Our dataset comprises nearly 17 years of \textit{Fermi}-LAT Pass 8 \citep{atwood2013_PASS8} R3 \citep{bruel2018_R3} observations, spanning from August 4, 2008, to May 1, 2025 and in the energy range from 0.1 to 100 GeV. The data analysis was performed with \texttt{easyfermi}\footnote{\url{https://github.com/ranieremenezes/easyFermi}} 2.0.18 \citep{demenezes2022easyfermi}, which is heavily based on \texttt{fermipy}\footnote{\url{https://fermipy.readthedocs.io/en/latest/index.html}} 1.4.0 \citep{wood2017fermipy} and \texttt{fermitools}\footnote{\url{https://fermi.gsfc.nasa.gov/ssc/data/analysis/software/}} 2.2.0, by means of a binned likelihood analysis and using NEWMINUIT as the minimizer.

The $15^{\circ} \times 15^{\circ}$ region of interest (RoI) was centered on BL~Lacertae and modeled with all 4FGL-DR4 sources \citep{abdollahi2020_4FGL,abdollahi2022_4FGL_DR3} plus additional point-like sources detected above $4\sigma$. The spectral shape and normalization of sources within $7^{\circ}$ of BL~Lacertae were left free to vary during the fit, while those up to $5^{\circ}$ outside the RoI were fixed in their 4FGL-DR4 values. For each energy decade, we divided the data into eight logarithmic bins, using SOURCE events (evclass = 128) detected in the front or back layers of the tracker (evtype = 3). We applied standard quality cuts (DATA\_QUAL $> 0$, LAT\_CONFIG == 1) and excluded events with zenith angles $\theta_z > 90^{\circ}$. The Galactic and isotropic backgrounds\footnote{\url{https://fermi.gsfc.nasa.gov/ssc/data/access/lat/BackgroundModels.html}} were modeled with \texttt{gll\_iem\_v07} and \texttt{iso\_P8R3\_SOURCE\_V3\_v1}, and the normalization of both components was left free to vary. Detection significance was estimated as $\sqrt{TS}$, where $TS = 2(\mathcal{L}_1 - \mathcal{L}_0)$ compares the maximum log-likelihoods with and without the target \citep{mattox1996_TS}.

We built an adaptive-binning photon flux light curve for BL~Lacertae (top panel of Fig.~\ref{fig:LCs_and_specs}) using the \texttt{easyfermi} method\footnote{\url{https://easyfermi.readthedocs.io/en/latest/lightcurve.html}} integrated from 0.1 to 100 GeV, in which we start by building a constant-binning light curve (20 bins in our case) and then further slice the bins with $TS > 2TS_{th}$ into smaller pieces. We set the TS threshold as $TS_{th} = 9000$, which is a good compromise between variability resolution and photon statistics above 10~GeV, necessary to constrain the spectral shape in time intervals where $E_p$ shifts to higher energies. The final light curve has 47 bins, 21 of which do not exceed the TS threshold; however, this does not compromise the analysis since these bins correspond to lower fluxes and lower values of $E_p$, for which the log-parabolic curvature remains well constrained. The model adopted to build the light curve is the log-parabola available in \texttt{fermitools} 2.2.0\footnote{\url{https://fermi.gsfc.nasa.gov/ssc/data/analysis/scitools/source_models.html}}, where we left the normalization, slope, and curvature free to vary.

For each bin in the adaptive-binning light curve, we computed a gamma-ray spectrum with the \texttt{fermipy} function \texttt{sed()}, corrected them for extragalactic background light absorption with \texttt{gammapy} \citep[rather small effect;][]{saldana2021_EBL,donath2023gammapy} adopting a redshift $z = 0.0686$ \citep{massaro2015romabzcat5th}, and used a Markov chain Monte Carlo \citep{Foreman-Mackey2013emcee} to estimate the parameters of the log-parabolic model \citep[see][]{massaro2004logpar2, tanihata2004evolution, tramacere2007signatures} conveniently defined by
\begin{equation}
    \label{eq:logparabola}
     S(E) = E^2dN/dE = S_p 10^{-\beta \log_{10}^2(E/E_p)},
\end{equation}
where $S_p$ is the differential energy flux at the log-parabola peak in erg cm$^{-2}$ s$^{-1}$, $\beta$ is the spectral curvature, and $E_p$ is the position of the log-parabola peak in the energy axis in MeV. This parametrization is convenient since we can directly estimate $S_p$ and $E_p$, as well as their corresponding errors, without resorting to huge error propagation formulas. The adopted priors span broad ranges so that they remain unchanged across all fits. Specifically, we use a flat prior distribution for $0 < \beta < 1$, and log-flat prior distributions for $100 < E_{p} < 10^{5}$ and $10^{-12} < S_{p} < 10^{-8}$. In the bottom panels of Fig. \ref{fig:LCs_and_specs}, we show three gamma-ray spectra of BL Lacertae selected at the three light curve bins tagged with arrows in the upper panel. From left to right, they have $TS = 8988, 8111, 12337$, and $E_p = 205^{+50}_{-59}, 724^{+100}_{-100}, 1343^{+275}_{-238}
$ MeV, respectively. Here we see that an increase in $S_p$ is followed by an increase in $E_p$. The dependency of $S_p$ on $E_p$ will reveal the physics causing the spectral variability, as discussed in the next sections.

\subsection{Multiwavelength data}
\label{subsec:SED_multiwavelength_data}

To estimate the parameters for the broadband SED of BL Lacertae, we selected 13 lower energy data points from archival observations using the online SED builder tool\footnote{\url{https://tools.ssdc.asi.it/}}. These observations and respective references are listed in Table \ref{tab:observations} in Appendix \ref{appendix:observation_dates}. We use these observations in the next section to roughly constrain the single-zone SSC parameters for BL Lacertae. A precise modeling of the SED is not the goal at this point of this work since we are interested only in the evolution of the SED and, in particular, in how $S_p$ grows with $E_p$. These archival data points range from radio to X-rays, while for the gamma-ray energies, we used the integrated \textit{Fermi}-LAT spectrum for the whole time window of almost 17 years.



\section{Results}
\label{sec:results}

\begin{figure}
    \centering
    \includegraphics[width=\linewidth]{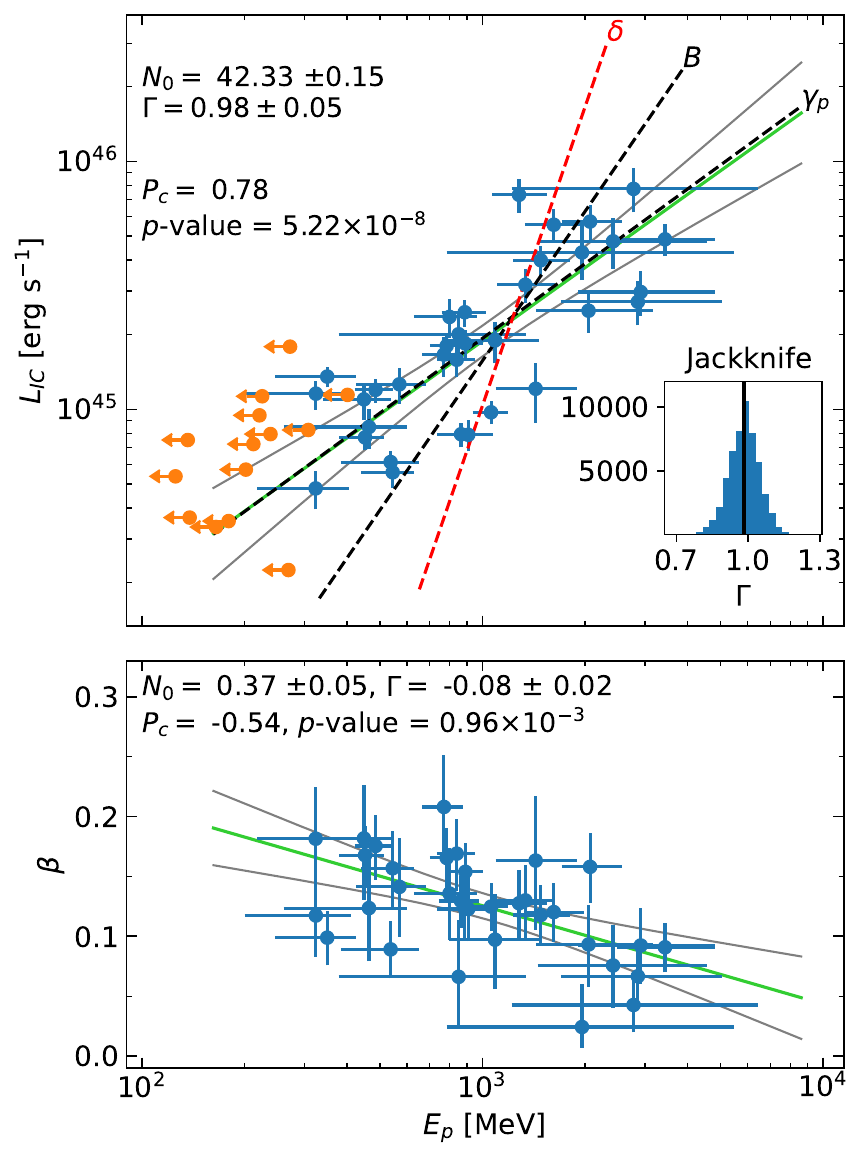}
    \caption{Rest-frame correlations of $L_{IC}$ and $\beta$ with $E_p$. Top: $L_{IC}$ scales nearly linearly with $E_p$, consistent with variability driven mainly by changes in $\gamma_p$. The inset shows the jackknife resampling distribution of the slope, centered at $\Gamma = 0.98$. Orange points mark upper limits on $E_p$ (not included in the fit), defined when $E_p < 300$ MeV or when uncertainties extend below 100 MeV. Dashed lines indicate the correlations expected if variability depends on a single parameter (see text). Bottom: The spectral curvature $\beta$ shows no significant dependence on $E_p$, suggesting BL~Lacertae may be transitioning out of the Thomson regime. $N_0$ and $\Gamma$ are described in the text.}
    \label{fig:correlation}
\end{figure}

The evolution of the isotropic SED peak luminosity ($L_{IC} = 4\pi D_L^2 S_p$, with $D_L$ the luminosity distance) as a function of $E_p$ is shown in the top panel of Fig.~\ref{fig:correlation}. The orthogonal distance regression (ODR) fit (green line), parameterized as $L_{IC} = 10^{N_0}E_p^{\Gamma}$, yields $N_0 = 42.33 \pm 0.15$ and $\Gamma = 0.98 \pm 0.05$, consistent within errors (gray lines) with a single-zone SSC model where only $\gamma_p$ varies. A jackknife resampling (inset) confirms the robustness of this correlation, giving 
\[
L_{IC} = 10^{42.33 \pm 0.15 \pm 0.18_{\rm sys}} E_p^{0.98 \pm 0.05 \pm 0.06_{\rm sys}},
\] 
where the systematic errors represent the $\sigma$ of the jackknife parameter distributions. This result is free of bias driven by up to four points in our dataset (representing more than 10\% of our sample). We also find a relatively strong Pearson correlation coefficient of $P_c = 0.78$ with a p-value for non-correlation of $5.2 \times 10^{-8}$, corresponding to $>5\sigma$ significance. For comparison, the black dashed lines in Fig.~\ref{fig:correlation} show the expected Thomson-regime scalings for variations in $\gamma_p$ ($\Gamma = 1$) or $B$ ($\Gamma = 2$), while the red dashed line represents the case where only the beaming factor $\delta$ varies, giving $\Gamma = 4$ in both Thomson and Klein–Nishina regimes.

The upper limits in Fig.~\ref{fig:correlation} are defined when $E_p < 300$~MeV or when uncertainties allow $E_p$ values below 100~MeV. Adopting alternative thresholds of 200~MeV or 400~MeV yields $L_{IC} = 10^{42.61 \pm 0.12 \pm 0.14_{\rm sys}} E_p^{0.90 \pm 0.05 \pm 0.05_{\rm sys}}$ and $L_{IC} = 10^{42.12 \pm 0.17 \pm 0.14_{\rm sys}} E_p^{1.06 \pm 0.06 \pm 0.05_{\rm sys}}$, respectively, both consistent with the linear correlation expected from variations in the electron peak energy.

For a log-parabolic electron distribution, the synchrotron curvature is expected to scale as $b \propto 1/\log(E_p)$ \citep{Tramacere2011jetset}. For the IC peak in the Thomson regime (as suggested for BL~Lacertae by the correlation in the top panel of Fig.~\ref{fig:correlation}), the curvature $\beta$ should follow a similar but systematically smaller trend due to energy redistribution in the scattering process \citep{massaro2006logpar3}. In general, $\beta$ is expected to anticorrelate with $E_p$ in the Thomson regime, show no correlation in the transition to the Klein–Nishina regime, and become positively correlated in the Klein–Nishina regime \citep{Tramacere2011jetset}. Our data (bottom panel of Fig.~\ref{fig:correlation}) shows a weak negative trend, $\beta = (0.37 \pm 0.05) E_p^{-0.08 \pm 0.02}$, with $P_c = -0.54$ and a p-value $\sim 10^{-2}$. These results suggest BL~Lacertae may be entering the hybrid Thomson/Klein–Nishina regime, where no clear $\beta$–$E_p$ correlation is expected. This trend also indicates that the highest luminosity SEDs will tend to have smaller curvatures, which are harder to constrain, explaining why some of our data points in Fig. \ref{fig:correlation} with luminosities above $\sim 3\times10^{35}$ erg~s$^{-1}$ present relatively large uncertainties in $E_p$.

\subsection{Precise SED evolution}
\label{subsec:SED_evolution}

\begin{figure*}
    \centering
    \includegraphics[width=\linewidth]{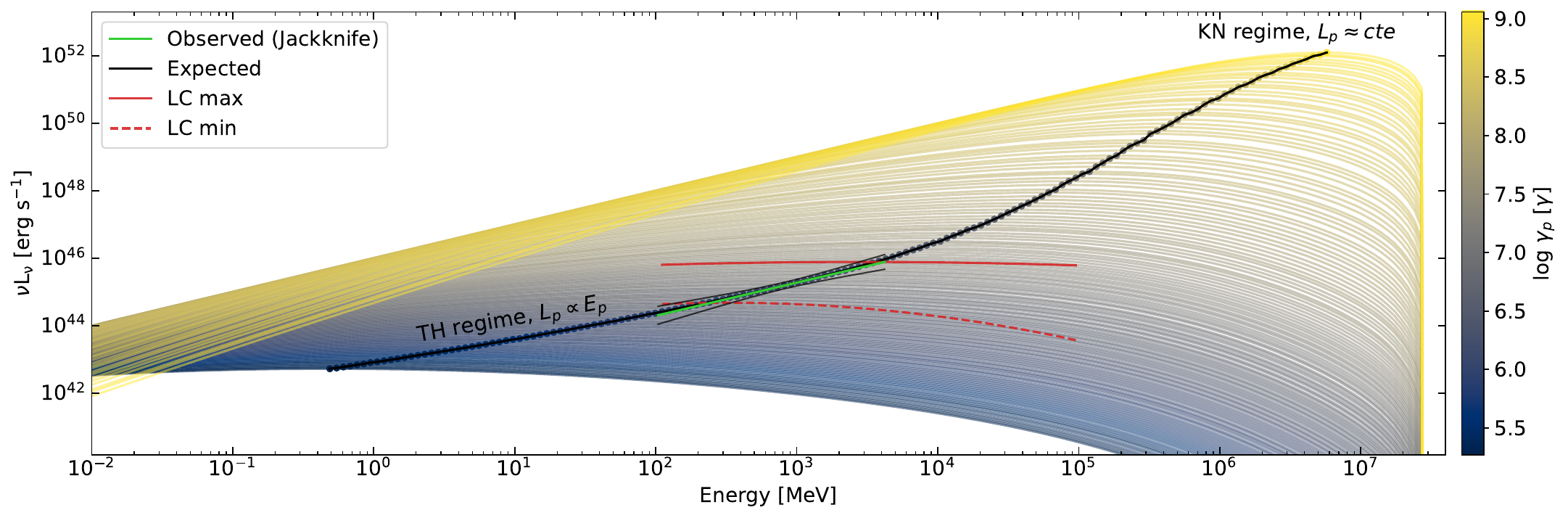}
    \caption{Evolution of BL~Lacertae's SED when only the electron peak energy $\gamma_{p}$ (or equivalently $\gamma_{0}$) is free to vary, while all other parameters are fixed to the values in Table~\ref{tab:starting_parameters}. The black line tracks the IC peak positions, the green line shows the jackknife correlation derived from the data, and the color gradient represents the SSC emission models for different values of $\gamma_{p}$. The results indicate that BL~Lacertae remains mostly in the Thomson regime, with hints of entering a Thomson/Klein-Nishina transition during its brightest flares. The red curves correspond to the log-parabolic models fitted to the highest- and lowest-flux bins shown in the light curve of Fig.~\ref{fig:LCs_and_specs}.}
    \label{fig:advanced_model}
\end{figure*}

Assuming a log-parabolic electron distribution (Sect.~\ref{sec:intro}) within a single-zone SSC framework, we use JetSeT 1.3.0\footnote{\url{https://jetset.readthedocs.io/en/1.2.1.post2/index.html}} \citep{Tramacere2009jetset,Tramacere2011jetset,Tramacere2020jetset} to study the evolution of BL~Lacertae's SED when only the peak electron energy $\gamma_{p}$ (or equivalently $\gamma_{0}$, see Sect. \ref{sec:intro}) is allowed to vary. To set up the model, we first fit the archival multiwavelength data described in Sect.~\ref{subsec:SED_multiwavelength_data}, and then adopt the rounded parameters of this fit as our starting model (see Fig.~\ref{fig:best_fit_SED} and Table~\ref{tab:starting_parameters} in Appendix~\ref{appendix:observation_dates}). 

With all other parameters fixed, we let $\gamma_p$ vary from $10^{5.5}$ to $10^9$ ($10^{3} \lesssim \gamma_{0} \lesssim 10^7$). The resulting SED evolution follows the expected trend: linear $L_p$--$E_p$ scaling in the Thomson regime, steeper growth in the intermediate Thomson/Klein-Nishina regime, and eventual convergence of $L_p$ to a constant at the highest $\gamma_{p}$ values (Fig.~\ref{fig:advanced_model}).  
In that figure, the blue-to-yellow gradient traces IC spectra as $\gamma_{p}$ increases, while the green line shows the jackknife correlation derived from our data. The results indicate that BL~Lacertae is mostly in the Thomson regime, though during strong flares it approaches the Thomson/Klein-Nishina transition, consistent with the weak $\beta$–$E_p$ correlation found earlier \citep[see Fig. 10 in][for a detailed discussion on this topic]{Tramacere2011jetset}.



\section{Discussion and conclusions}
\label{sec:discussion_and_conclusions}


In this work, we argue that the temporal and spectral variability of BL~Lacertae in $\gamma$ rays can be explained mainly by changes in the electron peak energy $\gamma_p$ in a single-zone SSC scenario. The expected correlation $L_{IC} \propto E_p$ is well reproduced by our result, $L_{IC} = 10^{42.33\pm0.15\pm0.18_{\rm sys}}E_p^{0.98\pm0.05\pm0.06_{\rm sys}}$, which is also in line with previous multiwavelength studies of blazars in different activity states \citep{dammando2013long,dutka2013multi,dutka2017multiband}. However, the relatively large scatter ($RMS_{\perp} = 0.19$ in the log space, which is comparable to the average effective orthogonal uncertainties, given by $\langle\sigma_{\perp,i}\rangle = \langle\sqrt{(\sigma_{x,i}\sin\theta)^2 + \sigma_{y,i}\cos\theta)^2}\rangle =  0.27$, where $\theta = \arctan(\Gamma)$) shown in Figure~\ref{fig:correlation} opens the possibility for other parameters to play a secondary role in the observed variability. Furthermore, the weak negative (or absent) correlation $\beta = (0.37\pm0.05)E_p^{-0.08\pm0.02}$ indicates that BL~Lacertae is likely approaching a hybrid Thomson/Klein-Nishina regime \citep[see][for a discussion on this topic]{vandenberg2019ApJ_systematic}.

In the Thomson regime, it is also possible to have a correlation of the form $L_{IC} \propto E_p^{\Gamma}$, with $\Gamma < 1$, and depending only on $\gamma_{p}$. This holds if we make the further assumption that the size of the emitting region, $R$, depends on the electrons' diffusion coefficient ($D$) and cooling time scale ($\tau$), $R = \sqrt{D\tau} \propto \gamma^{(\alpha-1)/2}$ \citep{haocheng2024revisiting}, where the index $\alpha$ can range from $1/3$ in the Kolmogorov turbulence regime to 1 in the Bohm approximation, and $\gamma$ is the energy of the electrons, here approximated as $\gamma \approx \gamma_{p}$, meaning that we assume a roughly mono-energetic electron population. For example, if $\alpha=1/3$, the expected correlation computed with JetSeT for BL Lacertae in the Thomson regime is $L_{IC} \propto E_p^{0.6}$, while if $\alpha = 0.85$, the expected correlation is $L_{IC} \propto E_p^{0.9}$, as found in our alternative analysis when we set the upper limits threshold at 200 MeV. In this last case, the dependency of $R$ on $\gamma_{p}$ is minimal, going as $R \propto \gamma_{p}^{-0.075}$. Finally, if $\alpha = 1$, we fall exactly in the correlation discussed in the rest of this work (i.e., $L_{IC} \propto E_p$), where the size of the emission region is a constant.

This work required a large amount of high-quality gamma-ray data, enabled by the exceptional brightness of BL~Lacertae, the location of its IC peak near 500~MeV, and the uninterrupted all-sky monitoring provided by Fermi-LAT. Other promising candidates for future studies include the BL~Lac objects Mrk~421 and S5~0716+71. However, their $\gamma$-ray spectra are often better described by a power law with an exponential cutoff (PLEC) rather than a log-parabola, meaning that a parametrization of the PLEC in terms of $S_p$ and $E_p$, such as the one available in easyfermi\footnote{\url{https://easyfermi.readthedocs.io/en/latest/SED.html}} \citep{demenezes2022easyfermi} may be useful:
\begin{equation}
    \label{eq:PLEC}
    S(E) = S_p\left( \frac{E_p}{E} \right)^{a - 2}e^{((2-a)/b) (1 - (E/E_p)^b)},
\end{equation}
where $a$ is the power law spectral index, and $b$ is the super-exponential index. Moreover, their IC peaks are located at higher energies, meaning that the usage of IACTs, such as the Large-Sized Telescope \citep{abe2023observations_LST} or the Cherenkov Telescope Array \citep{bernlohr2013_CTA}, may be necessary. Extending this analysis to other BL Lacs in the gamma-ray band would yield further insights into their emission mechanisms, particularly if their variability also proves to be primarily governed by the single parameter $\gamma_p$.


\begin{acknowledgments}
We would like to thank Justin Finke, Filippo D'Ammando, and the anonymous referee for the valuable comments and suggestions, which have substantially improved this manuscript. Part of this work is based on archival data and online services provided by the Space Science Data Center - ASI, especially with the SED builder tool\footnote{\url{https://tools.ssdc.asi.it/}}. HZ is supported by NASA under award number 80GSFC24M0006.

The \textit{Fermi} LAT Collaboration acknowledges generous ongoing support from a number of agencies and institutes that have supported both the development and the operation of the LAT as well as scientific data analysis. These include the National Aeronautics and Space Administration and the Department of Energy in the United States, the Commissariat \`a l'Energie Atomique and the Centre National de la Recherche Scientifique / Institut National de Physique Nucl\'eaire et de Physique des Particules in France, the Agenzia Spaziale Italiana and the Istituto Nazionale di Fisica Nucleare in Italy, the Ministry of Education, Culture, Sports, Science and Technology (MEXT), High Energy Accelerator Research Organization (KEK) and Japan Aerospace Exploration Agency (JAXA) in Japan, and the K.~A.~Wallenberg Foundation, the Swedish Research Council and the Swedish National Space Board in Sweden. Additional support for science analysis during the operations phase is gratefully acknowledged from the Istituto Nazionale di Astrofisica in Italy and the Centre National d'\'Etudes Spatiales in France. This work performed in part under DOE Contract DE-AC02-76SF00515.
\end{acknowledgments}

\appendix

\section{SED multiwavelength data and parameters}
\label{appendix:observation_dates}

\begin{figure*}
    \centering
    \includegraphics[width=\linewidth]{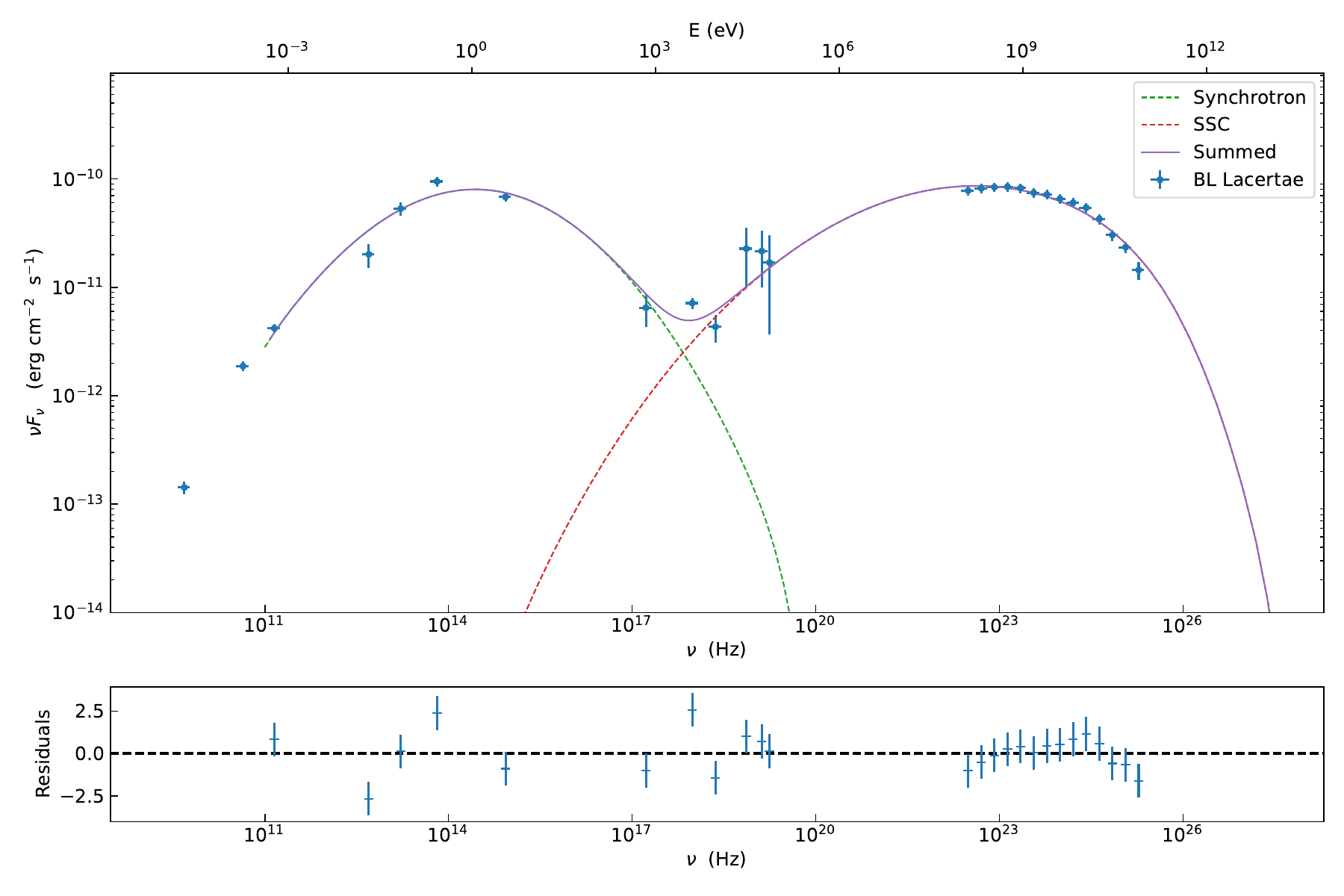}
    \caption{The best fit one-zone SSC model for the multiwavelength data described in Sect. \ref{subsec:SED_multiwavelength_data}. The rounded parameters for this fit are found in Table \ref{tab:starting_parameters}. We use these parameters as a starting point in the investigation of the SED spectral evolution when $\gamma_{p}$ (or, equivalently here, $\gamma_0$) is the only free variable. The \textit{Fermi}-LAT data shown here covers the 17 years of observations.}
    \label{fig:best_fit_SED}
\end{figure*}

In Table~\ref{tab:observations}, we list the multiwavelength archival and new observations used to model the SED of BL~Lacertae. The corresponding SSC model rounded parameters are given in Table~\ref{tab:starting_parameters}, while uncertainties are omitted because we are interested only on approximate values. The spectral slope of the electron distribution was set as $s = 2.6$ for convenience, and has no impact on the results. The data and best-fit model from these tables are shown in Fig.~\ref{fig:best_fit_SED}.

\begin{table}
    \centering
    \begin{tabular}{l|c|c}
        Freq. ($\mathrm{Hz}$) & Flux ($\mathrm{erg\,cm^{-2}\,s^{-1}}$) & Reference\\
        \hline
        $4.85 \times 10^{9}$  & $(1.43\pm0.13)\times10^{-13}$ & \cite{gregory1996gb6}\\  
        $4.40 \times 10^{10}$ & $(1.87\pm0.07)\times10^{-12}$ & \cite{ade2011planckERCSC}\\  
        $1.43\times10^{11}$  & $(4.19\pm0.09)\times10^{-12}$ & \cite{ade2011planckERCSC}\\
        $5.00\times10^{12}$  & $(2.01\pm0.46)\times10^{-11}$ & \cite{wang2009imperial}\\
        $1.67\times10^{13}$  & $(5.32\pm0.57)\times10^{-11}$ & \cite{ishihara2010akari}\\
        $6.52\times10^{13}$  & $(9.47\pm0.22)\times10^{-11}$ & \cite{cutri2021vizier}\\
        $8.56\times10^{14}$  & $(6.82\pm0.04)\times10^{-11}$ & \cite{yershov2014serendipitous}\\
        $1.69\times10^{17}$  & $(6.49\pm2.06)\times10^{-12}$ & \cite{giommi2002bepposax}\\
        $9.63\times10^{17}$  & $(7.14\pm0.46)\times10^{-12}$ & \cite{giommi2002bepposax}\\
        $2.30\times10^{18}$  & $(4.32\pm1.15)\times10^{-12}$ & \cite{giommi2002bepposax}\\
        $7.35\times10^{18}$  & $(2.27\pm1.23)\times10^{-11}$ & \cite{giommi2002bepposax}\\
        $1.31\times10^{19}$  & $(2.16\pm1.14)\times10^{-11}$ & \cite{giommi2002bepposax}\\
        $1.76\times10^{19}$  & $(1.69\pm1.31)\times10^{-11}$ & \cite{giommi2002bepposax}\\
        $3.09\times10^{22}$  & $(7.79\pm0.07)\times10^{-11}$ & This work\\
        $5.07\times10^{22}$  & $(8.18\pm0.05)\times10^{-11}$ & This work\\
        $8.30\times10^{22}$  & $(8.39\pm0.07)\times10^{-11}$ & This work\\
        $1.36\times10^{23}$  & $(8.46\pm0.07)\times10^{-11}$ & This work\\
        $2.23\times10^{23}$  & $(8.22\pm0.07)\times10^{-11}$ & This work\\
        $3.65\times10^{23}$  & $(7.44\pm0.06)\times10^{-11}$ & This work\\
        $5.97\times10^{23}$  & $(7.17\pm0.10)\times10^{-11}$ & This work\\
        $9.79\times10^{23}$  & $(6.55\pm0.12)\times10^{-11}$ & This work\\
        $1.60\times10^{24}$  & $(6.02\pm0.10)\times10^{-11}$ & This work\\
        $2.62\times10^{24}$  & $(5.43\pm0.18)\times10^{-11}$ & This work\\
        $4.30\times10^{24}$  & $(4.27\pm0.19)\times10^{-11}$ & This work\\
        $7.04\times10^{24}$  & $(3.03\pm0.20)\times10^{-11}$ & This work\\
        $1.15\times10^{25}$  & $(2.34\pm0.16)\times10^{-11}$ & This work\\
        $1.89\times10^{25}$  & $(1.44\pm0.24)\times10^{-11}$ & This work\\
    \end{tabular}
    \caption{Multiwavelength archival and new observations of BL~Lacertae. The \emph{Fermi}-LAT data listed here correspond to 17 years of observations.}
    \label{tab:observations}
\end{table}

\begin{table}
    \centering
    \begin{tabular}{l|c|c|c}
       Parameter  &  Value  &  Units & Free\\
       \hline
       Region size, $R$  &  $2.84\times 10^{16}$  & cm & Yes    \\
       Magnetic field, $B$  &  0.03   &  G & Yes \\
       Beaming factor, $\delta$ &  31.36  & Lor. factor & Yes \\
       Low-energy cutoff, $\gamma_{min}$ &  81.89  & Lor. factor  & Yes\\
       High-energy cutoff, $\gamma_{max}$ & $1.92\times10^{6}$ & Lor. factor & Yes\\
       Density of emitters, $n_0$ & 20.0 & cm$^{-3}$ & Yes  \\
       Turn-over energy, $\gamma_0$  &  $4.55\times10^3$   &  Lor. factor & Yes \\
       Spectral slope, $s$  &  2.6  & -- & No  \\
       Spectral curvature, $r$ & 0.58 & -- & Yes  
    \end{tabular}
    \caption{List of rounded SED parameters for BL Lacertae. The errors are omitted because they are not useful in our analysis.}
    \label{tab:starting_parameters}
\end{table}

\bibliography{apssamp}

\end{document}